\documentclass[a4paper,11pt]{article}
\usepackage{jinstpub} 
\usepackage{lineno}
\usepackage{amsmath}
\usepackage{multirow}

\title{\boldmath Status of Dual-Readout Calorimetry for Future High-Energy Physics Experiments}

\author[]{A. Pareti}
\collaboration[c]{on behalf of the IDEA dual-readout calorimeter group}
\affiliation{INFN and Università degli Studi di Pavia,\\
Via Bassi 6, Italy}

\emailAdd{andrea.pareti@pv.infn.it}

\abstract{Future experiments at high energy $e^+e^-$ colliders will focus on extremely precise Standard Model measurements. Among the most important physics benchmarks, there is the capability to resolve the Higgs decays into W or Z pairs, in their completely hadronic decay modes (4 jets in the final state), only based on the invariant mass of the jet pair coming from decay of the on-shell boson. This translates into a relative energy resolution target of $30\%/\sqrt{E}$, well beyond current detector performances. Dual-readout calorimetry is a technique which aims to improve the energy resolution, for single hadrons and hadronic jets, exploiting the information produced by two different physical processes, namely scintillation and \v Cerenkov light emission. The IDEA detector, whose concept has been included in both the FCC and CEPC Conceptual Design Reports, is based on a dual-readout fibre calorimeter with independent fibre readout exploiting Silicon PhotoMultipliers (SiPMs). The individual SiPM information will be beneficial for a highly granular calorimeter design, opening up to advanced reconstruction techniques such as Particle Flow and a variety of neural network algorithms.
In this paper the status of calorimeter prototypes that have been developed to demonstrate the feasibility of the dual-readout method in association with the high granularity feature is illustrated. The specific choice for the design of each prototype is presented, together with the performances achieved at high-energy test beams or through simulations. }


\begin{document}
\maketitle
\flushbottom

\section{Physics requirements on future calorimeters}
\label{sec:physics_motivation}
The discovery of a Higgs boson-like particle by the ATLAS and CMS experiments in 2012 provided a definitive confirmation of the Standard Model predictive power. Future high-energy electron-positron colliders, such as the Future Circular Collider~\cite{Benedikt:2651299} or the Circular Electron-Positron Collider~\cite{thecepcstudygroup2018cepc}, will take advantage of the precise knowledge of the energy in the centre-of-mass frame and cleaner environment with respect to the LHC to reduce the systematic uncertainties and be able to identify also slight divergences from the Standard Model calculations. 

\begin{table}[h]
\centering
\caption{Working points at the FCC-ee.}
\label{tab:FCC-plan}
\smallskip

\resizebox{\textwidth}{!}{  \begin{tabular}{c|c|c|c|c|c|c}
        \hline
        Working Point                                        & Z, years 1-2    & Z, later                     & $W^+W^-$ & HZ &  \multicolumn{2}{c}{ $t\overline{t}$} \\ 
        \hline
        $\sqrt{s}$ (GeV)                                                                        &      \multicolumn{2}{c|}{88, 91, 94}                         &   157, 163 & 240                                     &   340-350 & 365                                 \\ 
        \hline
        Lumi/IP ($10^{34}$cm$^-{2}$s$^{-1}$) & 115 & 230 & 28 & 8.5 & 0.95 & 1.55 \\
        \hline
        Lumi/year (ab$^{-1}$, 2IP) & 24 & 48 & 6 & 1.7 & 0.2 & 0.34 \\
        \hline 
        Physics Goal (ab$^{-1}$) & \multicolumn{2}{c|}{150} & 10 & 5 & 0.2 & 1.5 \\
        \hline 
        Run time (year) & 2 & 2 & 2 & 3 & 1 & 4 \\
        \hline
         & \multicolumn{2}{c|}{}  &  &  $10^6$ HZ & \multicolumn{2}{c}{$10^6$ $t\overline{t}$} \\  
         
        Number of Events & \multicolumn{2}{c|}{$5\times 10^{12}$ Z}  & $10^8$ WW &  + & \multicolumn{2}{c}{+ 200k HZ} \\  
        
         & \multicolumn{2}{c|}{}  &  &  25k $W^+W^- \rightarrow H$ & \multicolumn{2}{c}{+50k $W^+W^- \rightarrow H$} \\  
        \hline 
     \end{tabular}}
\end{table}

Table~\ref{tab:FCC-plan} lists the energy working points that will be explored, the target luminosities and the expected number of physics events at the FCC-ee. An extended electroweak, Higgs and top quark physics precision-measurement programme is planned. Future $e^+e^-$ colliders will provide copious amounts of $b$, $c$ quarks and $\tau$ leptons, allowing for heavy-flavour physics studies. Both direct and indirect searches for Beyond-the-Standard-Model processes will also be conducted over the whole physics programme of future accelerators.
At all mentioned working points, the large hadronic branching fraction of the $W^{\pm}$, the $Z$ or the $H$ bosons leads to a high multiplicity of hadron jets in most interesting final states. 90\% of events will contain at least one hadronic jet. 
The capability to separate the on-shell W and Z bosons based on the invariant mass of the two consequent hadron jets will drastically increase the statistics available for the analyses, which is today mostly limited to leptonic decays. 
In order to have distinct high-purity samples of Higgs decays into W or Z boson pairs, the jet energy resolution is  required to be as good as  $3-4\%$ to statistically separate the invariant mass of  the on-shell W and Z boson.
As the calorimeter performance on energy relative resolution usually scales with $\sigma/E \propto 1\sqrt{E}$, this benchmark is usually translated to a target relative resolution for future calorimeters of $ \sigma/E = 30\%/\sqrt{E} $. This is a difficult requirement on hadron calorimetry and has currently never been achieved on a running collider experiment.  There are currently two main lines of research to achieve such a performance: highly granular calorimeters optimized for Particle Flow techniques, and dual-readout calorimeters. The first one is here briefly described, while the second is better detailed in Section~\ref{sec:DR_principles}. The idea behind Particle Flow calorimetry is to exploit as much as possible the information coming from the tracker detector, measuring with a much better precision the momenta of charged tracks through their bending inside a magnetic field. This technique greatly benefits from a very finely segmented calorimeter in both lateral and longitudinal directions, removing energy contributions coming from charged tracks and replacing them with the corresponding momenta. The remaining energy deposits in the calorimeter then belong to photons and other neutral particles, whose energy usually sum up to about$~10\%$ of the total jet energy~\cite{THOMSON200925}.

\section{Dual-readout calorimetry principles}
\label{sec:DR_principles}

Any hadron shower inherently comprises electromagnetic showers (through the $\pi^0, \eta \rightarrow 2\gamma$ decays) whose fraction of energy with respect to the total shower one is called electromagnetic fraction $f_{em}$. Naming $e$ the calorimeter response to the electromagnetic component of a shower and $h$ the non-electromagnetic one, calorimeters are tipically characterised by a ratio $e/h \neq 1$, property known as \emph{non compensation}. The limits on hadron energy measurements are related to large and event-based fluctuations in $f_{em}$, which exhibits a non-Gaussian and asymmetrical distribution, and that is also observed to increase with the overall shower energy (non-linearity of the detector response). 
The \emph{dual-readout method}~\cite{Lee_2018} aims at improving the energy resolution specifically on hadron showers, by measuring the $f_{em}$ event-by-event and using it to correct the reconstructed energy. To do so, signals produced by means of two physical processes emitted by active materials featuring different $e/h$ properties would be used. At present, all dual-readout projects focus on using \v Cerenkov light emission as one of these two processes: electrons and positrons, which are part of the electromagnetic component, represent the largest contributions of relativistic particles in hadron showers up to energies as low as 200~keV. Scintillation light emission is the other default choice since the number of photons emitted depends on the total number of ionising particles. After a calibration with an electron beam is performed, the signals obtained through these two processes are

\begin{equation}
\begin{aligned}
 S & = E [ f_{em} + (h/e)_S (1-f_{em}) ], \\
 C & = E [ f_{em} + (h/e)_C (1-f_{em}) ] 
\end{aligned}
\end{equation}
for scintillation and \v Cerenkov light, respectively. 
The dual-readout corrected energy measurement, obtained by means of the S and C signals, is then:
\begin{equation}
    E = \frac{S-\chi C}{1-\chi}, \qquad \text{with} \quad \chi = \frac{1-(h/e)_S}{1-(h/e)_C}.
\end{equation}
The two response ratios $(h/e)_S$ and $(h/e)_C$ are constant and, in principle, measurable quantities, and they are a measurement of non-compensation in the response of the two independent active materials. The benefits of this method are that the $\chi$ factor only depends on the design of the calorimeter and not on the nature of the incoming particle, guaranteeing a linear response (independent on fluctuations in the $f_{em}$) with a Gaussian distribution centred around the correct energy. The resulting dual-readout  calorimeter is then automatically compensating, featuring the same response for electromagnetic and hadronic showers.
It should be noticed that the $\chi$ parameter can be directly estimated at a test beam from the relation:
\begin{equation}
    \chi = \frac{E-S}{E-C}.
\end{equation}



\section{The IDEA dual-readout calorimeter}
The validity of the dual-readout calorimetry technique has been proved in over two decades of R\&D by the DREAM/RD52 collaborations~\cite{Wigmans_2012}\cite{GAUDIO2011339}, with different prototypes that have been built and tested using high-energy particle beams. Because of the promising results achieved by these groups, a dual-readout calorimeter has been considered to be part of the Innovative Detector for Electron-positron Accelerator (IDEA) detector for future $e^+e^-$ colliders.

\begin{figure}
    \centering
    \includegraphics[width=0.49\textwidth]{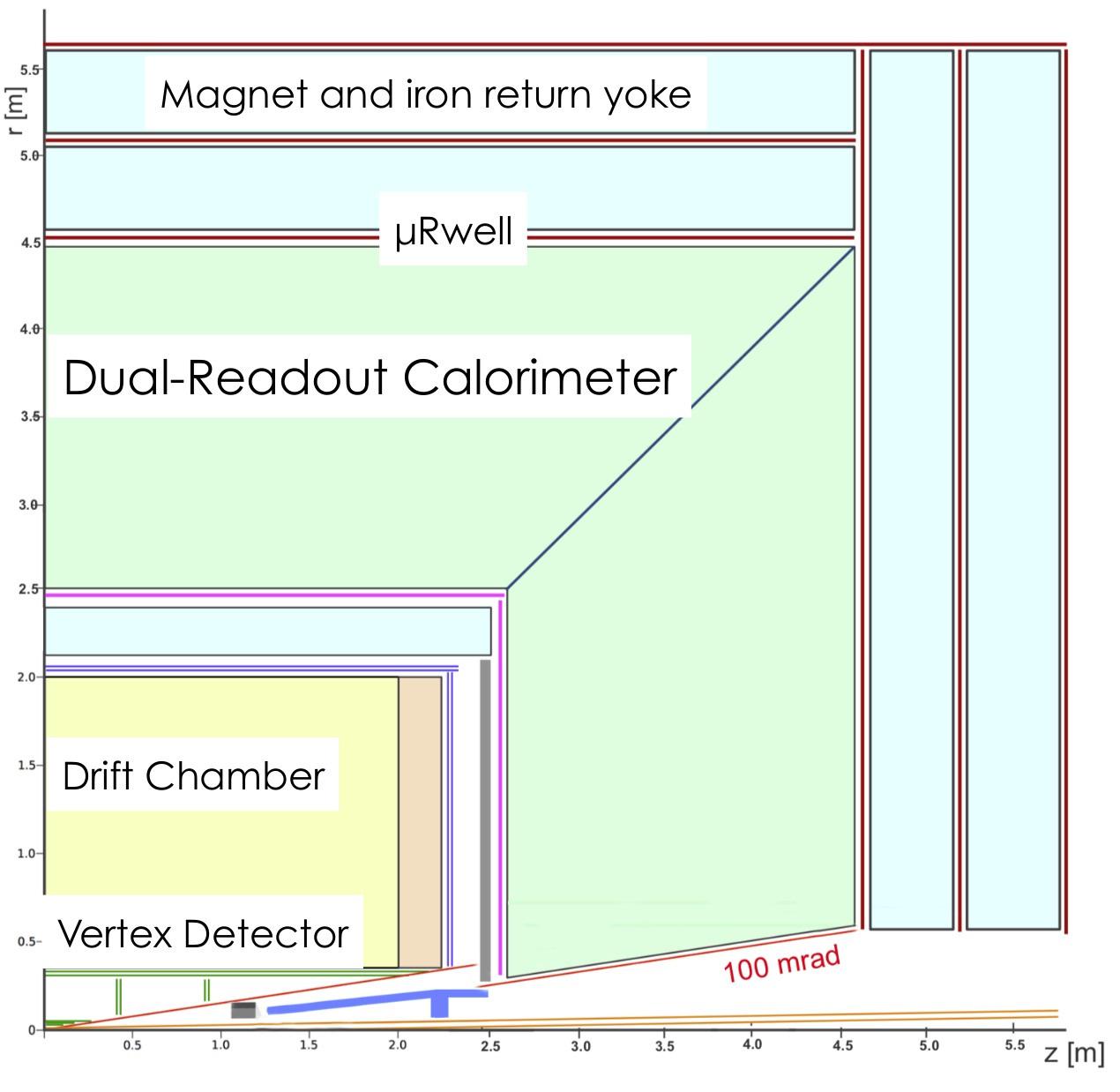}
    \hfill
    \includegraphics[width=0.49\textwidth]{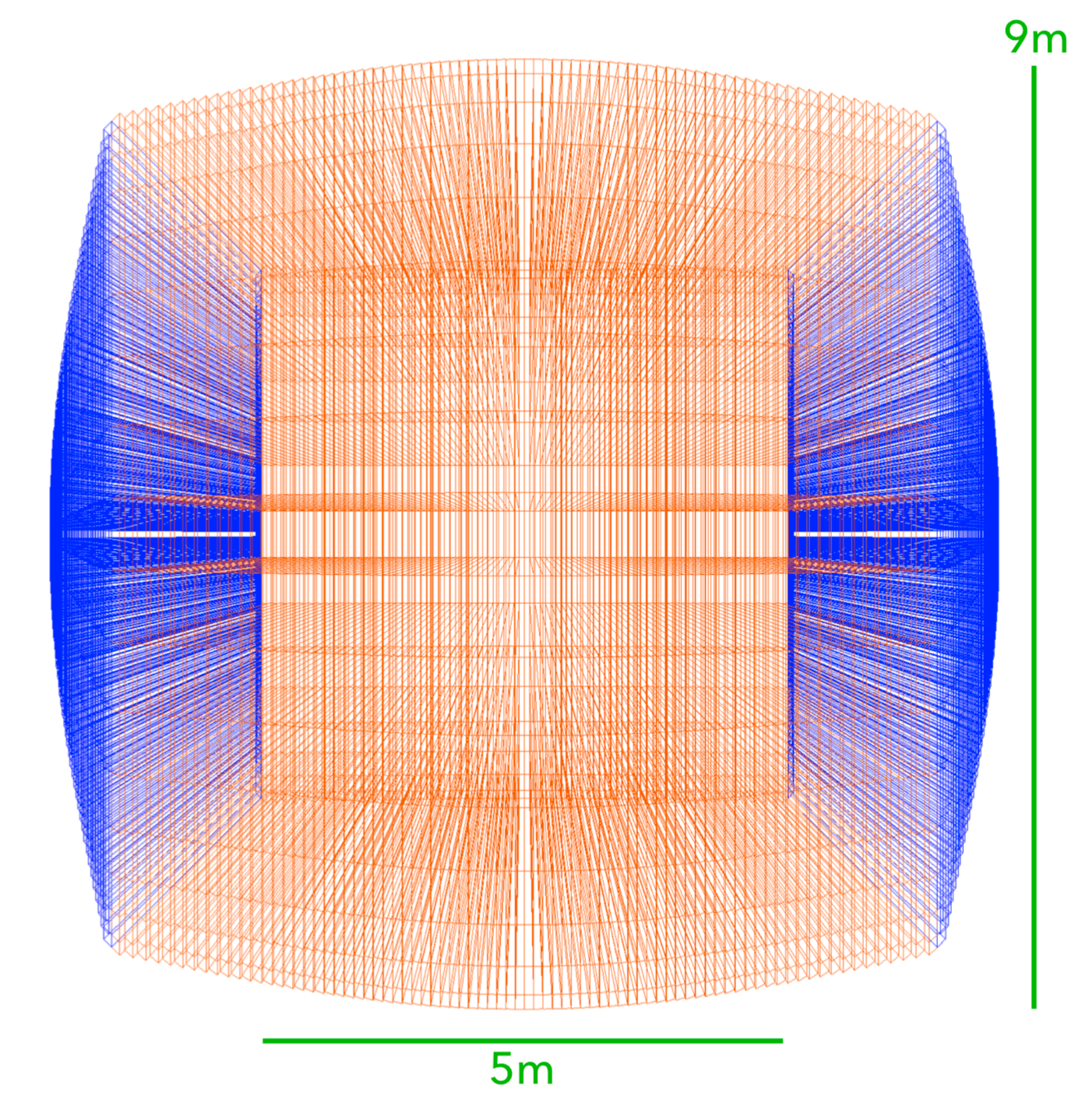}
    \caption{Sketch of the IDEA detector and Geant4 simulation of its dual-readout fibre calorimeter.}
    \label{fig:idea}
\end{figure}

IDEA has been included in both the FCC and CEPC Conceptual Design Reports as one of the experiments located in the collider interaction points. A sketch of its design is shown in Figure~\ref{fig:idea}: it features a silicon-based pixel detector surrounded by an ultralight drift chamber for the measurement of track momenta. The tracking system is immersed in a 2T magnetic field provided by a very thin solenoidal coil located before the dual-readout calorimeter volume. A Micro-Pattern Gaseous Detector (MPGD) called $\mu$-RWELL would be used for both the preshower detector and the muon system, immediately before and after the calorimeter respectively. The baseline solution for the dual-readout calorimeter consists of a longitudinally unsegmented fibre-based design, with fibres running almost parallel to the radial direction. A different design~\cite{Lucchini:2022goz} has also been proposed more recently, and not discussed here, consisting of a crystal-based dual-readout calorimeter section, whose purpose is further improving the resolution on electromagnetic showers at a level of $2-3\%/\sqrt{E}$. In both cases, the default design for hadron calorimetry consists of two different types of active plastic fibres, clear for \v Cerenkov light and doped for scintillation, inserted in capillary tubes of a common absorber material. The other important feature of this calorimeter is its very fine lateral segmentation, which is reached using Silicon PhotoMultipliers (SiPMs) to collect the photons that reach the rear end of each fibre. Among the main advantages are the high granularity of the resulting calorimeter, making it feasible for the usage of Particle-Flow algorithms, and the  removal of optical fibres sticking out from the back end of the absorber structure for the connection with classic PMTs.
SiPMs are fast photodetectors. Exploiting the time of arrival information, or other timing properties, with dedicated readout electronics will allow to obtain longitudinal position information  which can be used to improve particle identification in the dual-readout calorimeter.
In order to demonstrate the viability of a highly-granular dual-readout calorimeter to be deployed at the IDEA detector, smaller-scale prototypes able to contain electromagnetic showers have been assembled and tested with particle beams. Another essential part of the project is the definition of a construction technique, able to scale from a prototype to a full azimuthal coverage detector with a projective geometry. For this reason the prototypes that are presented in chapters \ref{sec:TB2021} and \ref{sec:hidra} feature a modular design consisting of small dimension units, able to be adapted at more complicated layouts.

\section{Electromagnetic-shower size prototype}
\label{sec:TB2021}

\begin{figure}
    \centering
    \includegraphics[width=.555\textwidth]{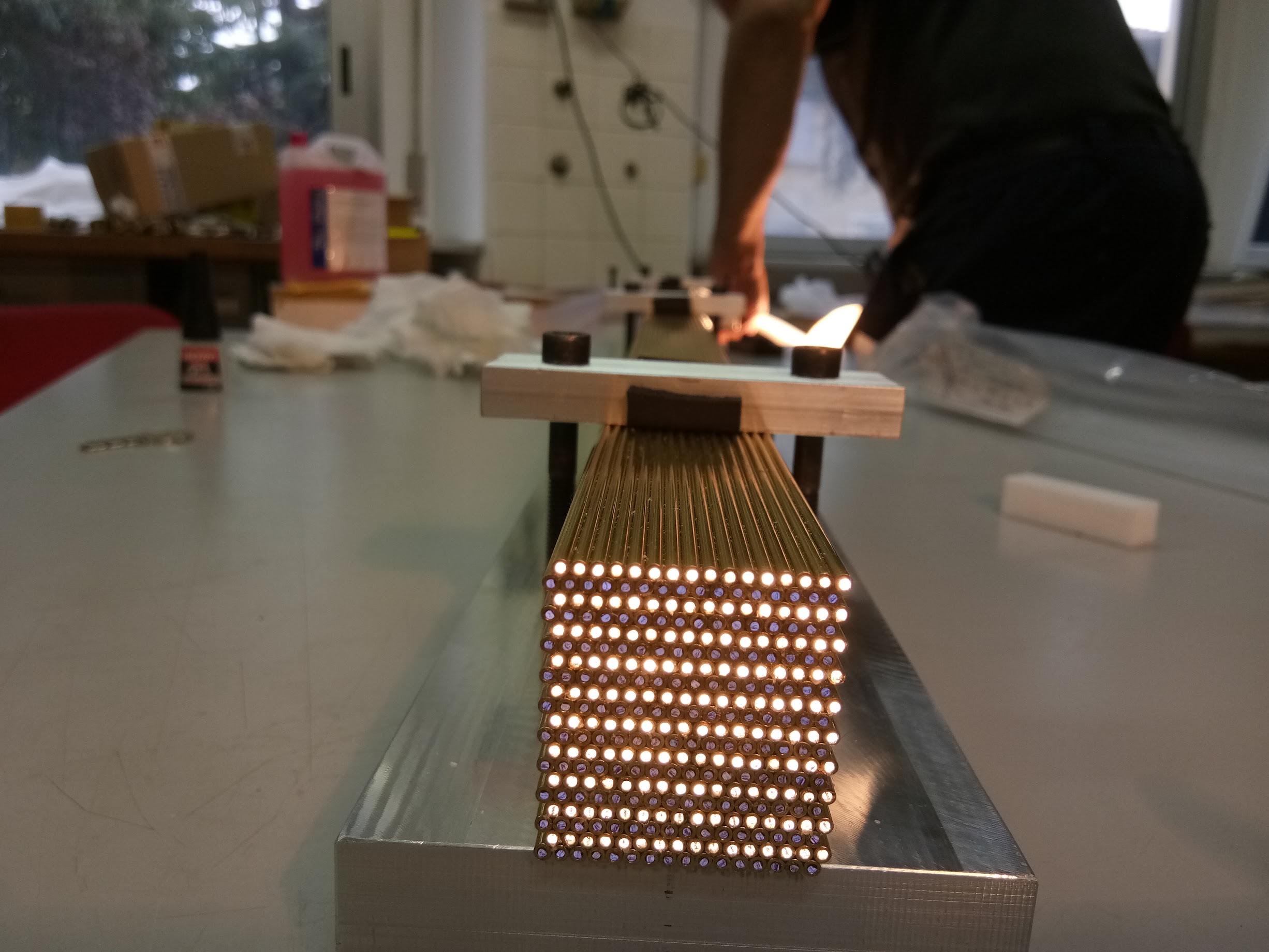}\hfill
    \includegraphics[width=.405\textwidth]{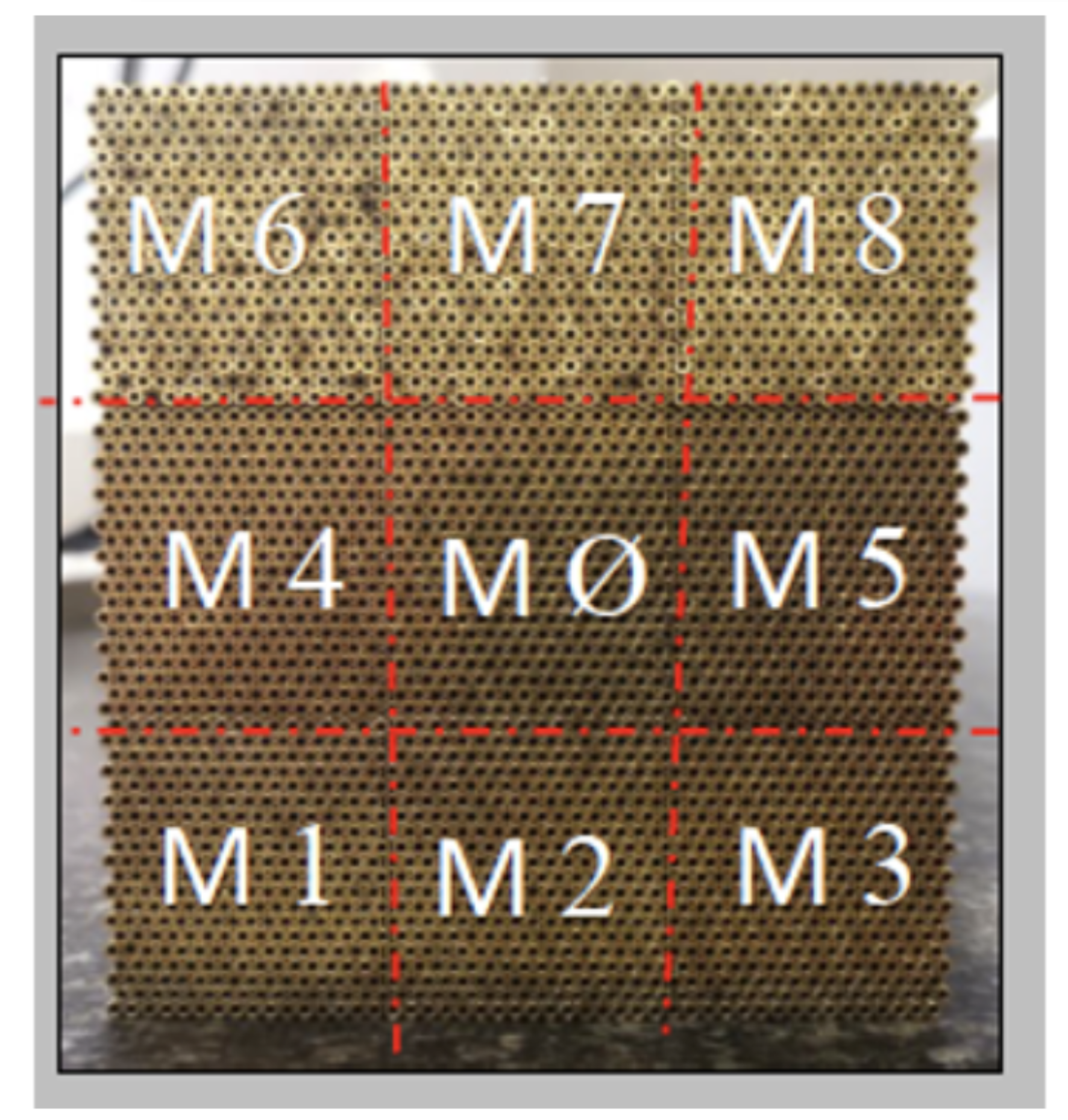}
    \caption{One module of the 2021 calorimeter prototype is shown on the left, with the \v Cerenkov clear fibres illuminated. Right image shows the front face of the full prototype, made of nine modules.}
    \label{fig:TB2021prototype}
\end{figure}

The first prototype of a dual-readout calorimeter prototype combined with SiPM readout sufficiently large to contain electromagnetic showers was built in 2021. It was then characterised at the DESY laboratories in Germany and at CERN's SPS~\cite{ampilogov2023exposing} with positron beams at different energies. The prototype geometry is shown in Figure~\ref{fig:TB2021prototype}: one module is made of 320 brass capillary tubes (63\% Cu, 37\% Zn) with an outer diameter of 2~mm and an inner one of 1.1~mm, arranged in 20 rows of 16 tubes each. Alternating rows of clear (Mitsubishi SK-40) and scintillating-doped (Saint-Gobain BCF-10) optical fibres, with an outer diameter of 1~mm, are inserted in the capillaries. The full prototypes consists of nine identical modules: the central one, called M0, is equipped with 320 S141601315 PS Hamamatsu SiPMs. Bundled scintillating fibres in the external modules (M1-M8) are equipped with Hamamatsu  R8900 PMTs, while the \v Cerenkov ones with the R8900-100 version for extended UV light range. A yellow filter is inserted at the back end of scintillating  fibres to cut off the shorter wavelength light component, such that the remaining signal (which is due to longer wavelengths) is less dependent on the shower starting point. The SiPM model was chosen for its wide dynamic range, using a 15 $\mu$m pitch and a sensitive area of $1.3\times1.3$~mm$^2$.  
The readout system for the SiPM equipped module is based on CAEN's FERS front-end board system, able to read up to 64 SiPMs per board. A total of five front-end boards were then used to read the highly-granular central tower.

\begin{figure}
    \centering
    \includegraphics[width=.48\textwidth]{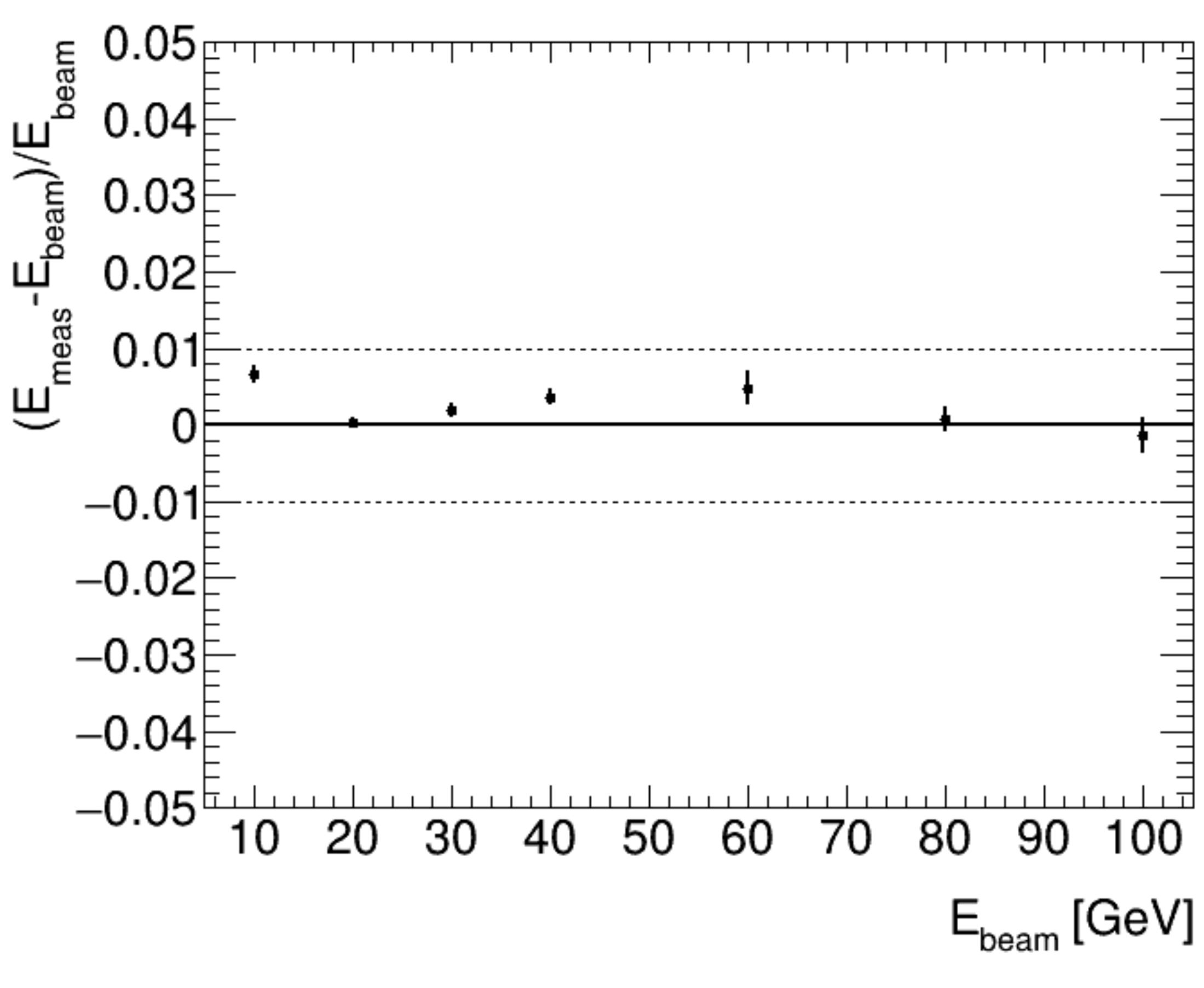}\hfill
    \includegraphics[width=.50\textwidth]{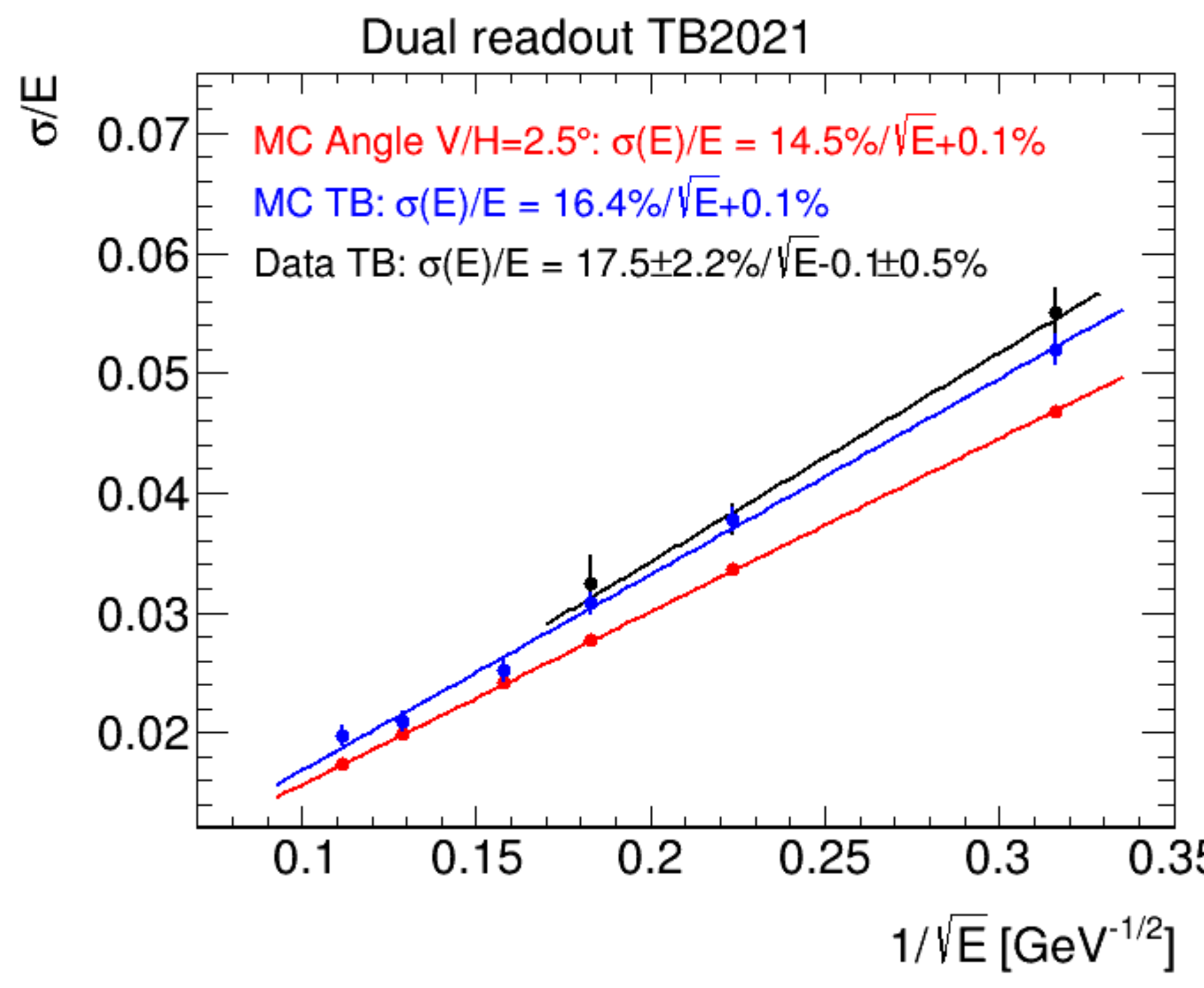}    
    \caption{Results from the electromagnetic prototype 2021 test beam campaign~\cite{ampilogov2023exposing}. The plot on the left shows the difference between the nominal beam energy and the calorimeter reconstructed one. On the right the relative resolution on energy measurement is plotted for the test beam data (black) and its Geant4 simulation (blue). The red line describe the expected behaviour given an inclination angle sufficient to remove the response modulation in the central tower rows.} 
    \label{fig:tb2021plots}
\end{figure}

A few results from the 2021 SPS test beam campaign, where the calorimeter was exposed to a positron beam with energies from 10 GeV up to 100 GeV, are presented in the following and shown in Figure~\ref{fig:tb2021plots}. As expected, the prototype has proved to be able to correctly reconstruct the true beam energy within a 1\% interval over the full energy range. For each energy measurements, the standard deviation and the mean value of a Gaussian fit were estimated. The energy resolution was calculated as the ratio of the two values. The resolution measurement was limited to only three energy working points due to a very large beam contamination of hadrons experienced during the test beam and the condition of auxiliary detectors located on the beam line. For energies lower than 30 GeV a reliable selection of electrons could be realized with \v Cerenkov threshold counters positioned on the beamline. On the other hand, at higher energies a preshower detector positioned upstream should have been used, but its positioning very far from the calorimeter due to access limitation led to a significant increase in the lateral leakage because of the limited prototype dimensions of $\sim10\times10\times100$ cm$^3$. Nonetheless, the results were used to validate a Geant4 simulation reproducing the test beam geometry, and the comparison showed a good agreement. The calibrated signal from the calorimeter was observed to suffer a modulation depending on the impact point position, with opposite phases between scintillating and \v Cerenkov fibres. It was observed through simulation studies that tilting the prototype with an angle of $2.5^{\circ}$ in both horizontal and vertical directions would have allowed to remove such effect almost completely. The energy measurement resolution obtained through the same simulation after a rotation on both X and Y axes with the mentioned angle was studied and reached $\sigma/E=14.5\%/\sqrt{E}+0.1\%$.

\begin{figure}
    \centering
    \includegraphics[width=.45\textwidth]{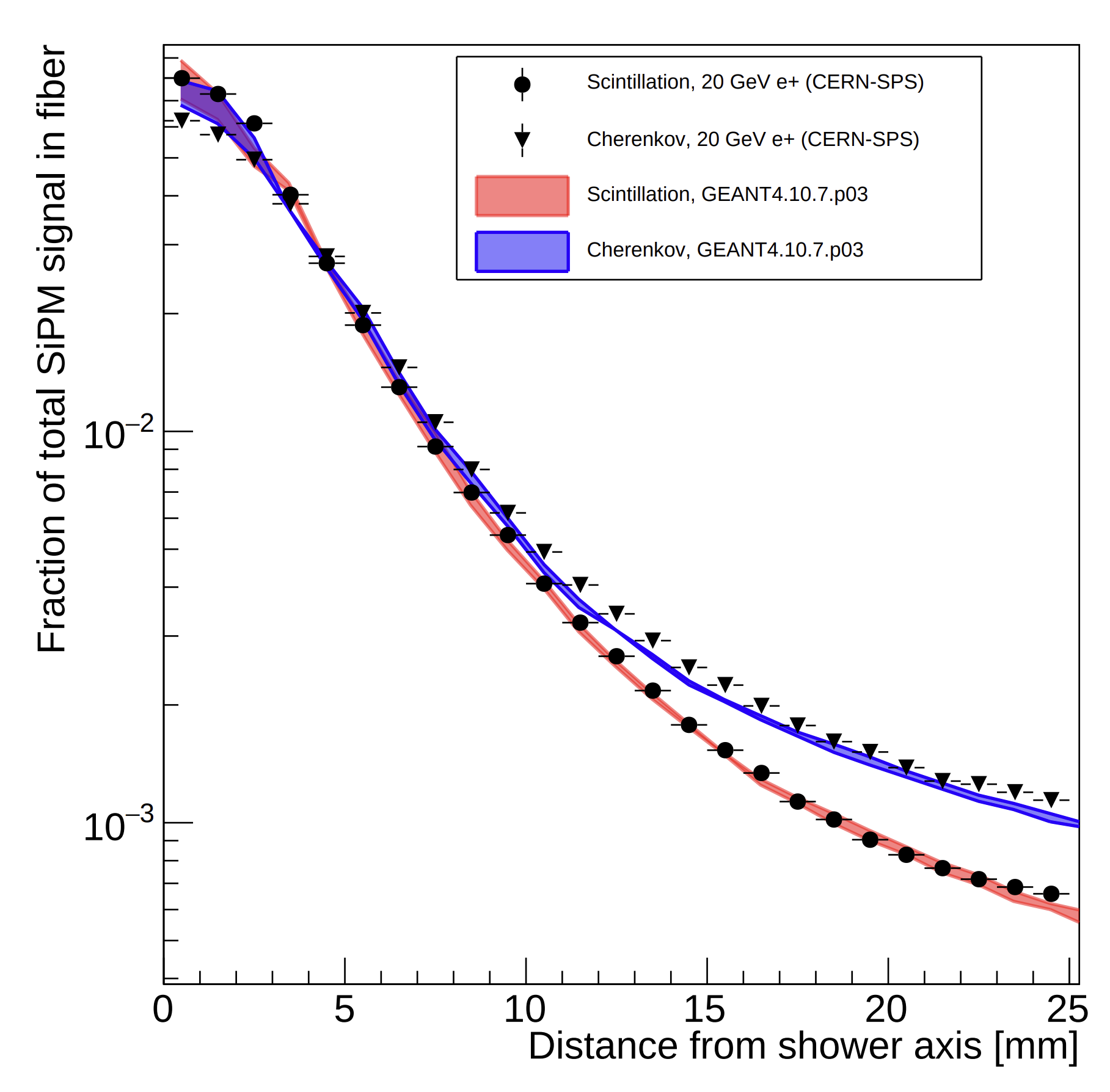}
    \caption{Lateral electromagnetic shower profile measured in the SiPM-equipped module~\cite{ampilogov2023exposing}. }
    \label{fig:showerprofile}
\end{figure}

The independent SiPM information allowed further studies on the electromagnetic shower development. Figure~\ref{fig:showerprofile} shows the lateral profile of showers produced by 20 GeV electrons in the highly granular tower of the prototype. The lateral profile is defined as the fraction of energy deposited in a fibre with respect to the distance from the shower axis. One finds that about 10\% of the total energy is deposited in less than two millimeters from the core of the shower. The shower development is found to be well described by the Geant4 simulation, with the scintillating and \v Cerenkov signals differing due to strong correlation, with the incoming positron direction, of the Čerenkov light emission in early development stages, while scintillation light is instead isotropic.

The same prototype has been taken to SPS North Area again in July 2023 for a further characterisation with a better beam detector layout. It has been exposed to a positron beam, whose purity was in a first instance estimated to be much improved with respect to the previous beam test one. Muons and pions beams were also used. At the time of writing, the analysis of the collected data is ongoing.

\section{Hadron shower-sized prototype: HiDRa}
\label{sec:hidra}

\begin{figure}
    \centering
    \includegraphics[width=.53\textwidth]{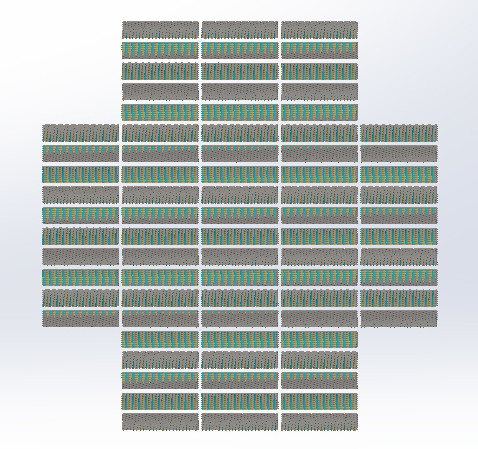}\hfill
    \includegraphics[width=.46\textwidth]{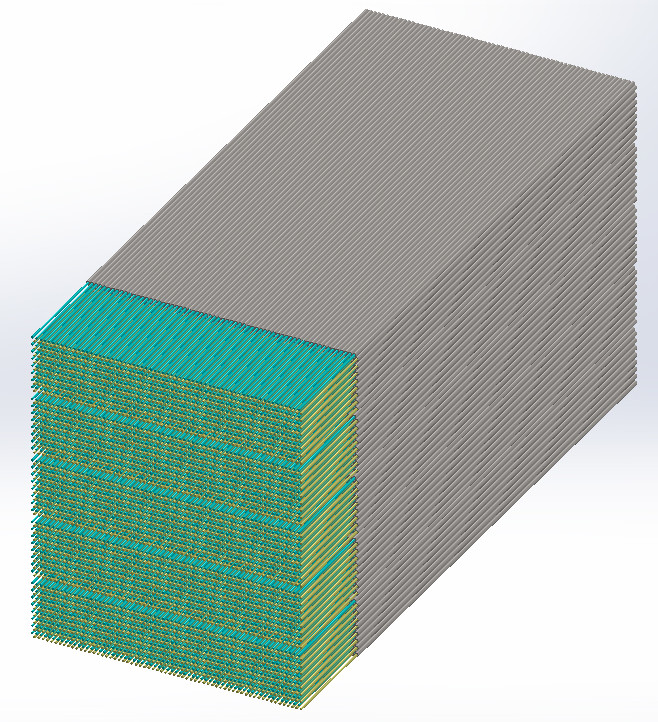}
    \caption{Front face of the HiDRa calorimeter prototype and one of its modules, made of five distinct minimodules. }
    \label{fig:hidradesign}
\end{figure}

The High-Resolution Highly-Granular Dual-Readout Demonstrator (HiDRa) is the second dual-readout calorimeter prototype associated to SiPM readout, and will be large enough to contain hadronic showers. It shares a similar blueprint with to the electromagnetic one, with the full prototype being composed of multiple identical  units called minimodules. The layout is illustrated in Figure~\ref{fig:hidradesign}. One minimodule accomodates 16 rows of 64 stainless steel capillary tubes each, with alternating rows of scintillating or clear undoped fibres. Five minimodules are aggregated into a full module. 
The two central modules, where the beam will be impinging, are going to be coupled to SiPMs, while all the others with two PMTs each, for scintillating and \v Cerenkov fibres respectively. This choice was determined by the need for a gradual increase in the amount of independent channels to be acquired by the readout system, and by the optimization of the expected performance over the available budget. 
The calorimeter design choices were guided by the same Geant4 simulation that was mentioned for the electromagnetic prototype and validated with positron beam, with minor changes on the geometry and by taking full advantage of the detector modularity. 

\begin{figure}
    \centering
    \includegraphics[width=\textwidth]{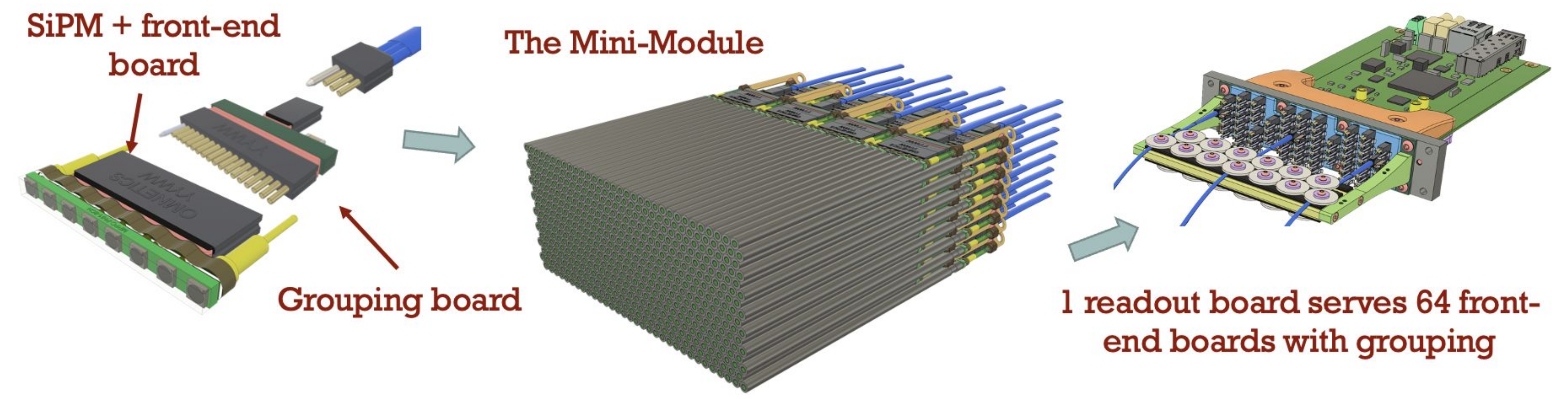}
    \caption{Front-end electronics connection to the HiDRa minimodules.}
    \label{fig:hidra_readout}
\end{figure}

Concerning the acquisition chain, two different SiPM models are being used, Hamamatsu S16676-15(ES1) for \v Cerenkov fibres and S16676-10(ES1) for scintillating ones. In the first case, to cope with the low light yield, a pixel pitch of 15 $\mu$m was selected for an improved photodetection efficiency, while, in the latter case, a pitch of 10 $\mu$m aiming at wider dynamic range is used. The CAEN A5202 FERS system, which exploits two WeeROC Citiroc 1A ASICs, is used as readout board for charge and timing measurements. Figure~\ref{fig:hidra_readout} shows how the front-end electronic boards are accustomed behind the limited volume of each minimodule. Great effort has been put on keeping the system as compact as possible, in order to fully demonstrate the capability of using this technology for a $4\pi$ coverage detector. To achieve this, and also reduce the number of channels and the related electronics costs, it was decided to have a grouping of signals from eight same-type fibres, in one single readout channel. The grouping board is directly located behind the SiPMs and builds the analogue sum of the signals that is then received by the readout board,  reaching a total of 64 channels per minimodule. Studies on the impact of this design choice on the spatial resolution with respect to single-fibre readout are at preliminary stages and will be deepened shortly, together with performance estimates of the high-granularity features exploiting particle-flow algorithms and neural network usage.

\begin{figure}
    \centering
    \includegraphics[width=.49\textwidth]{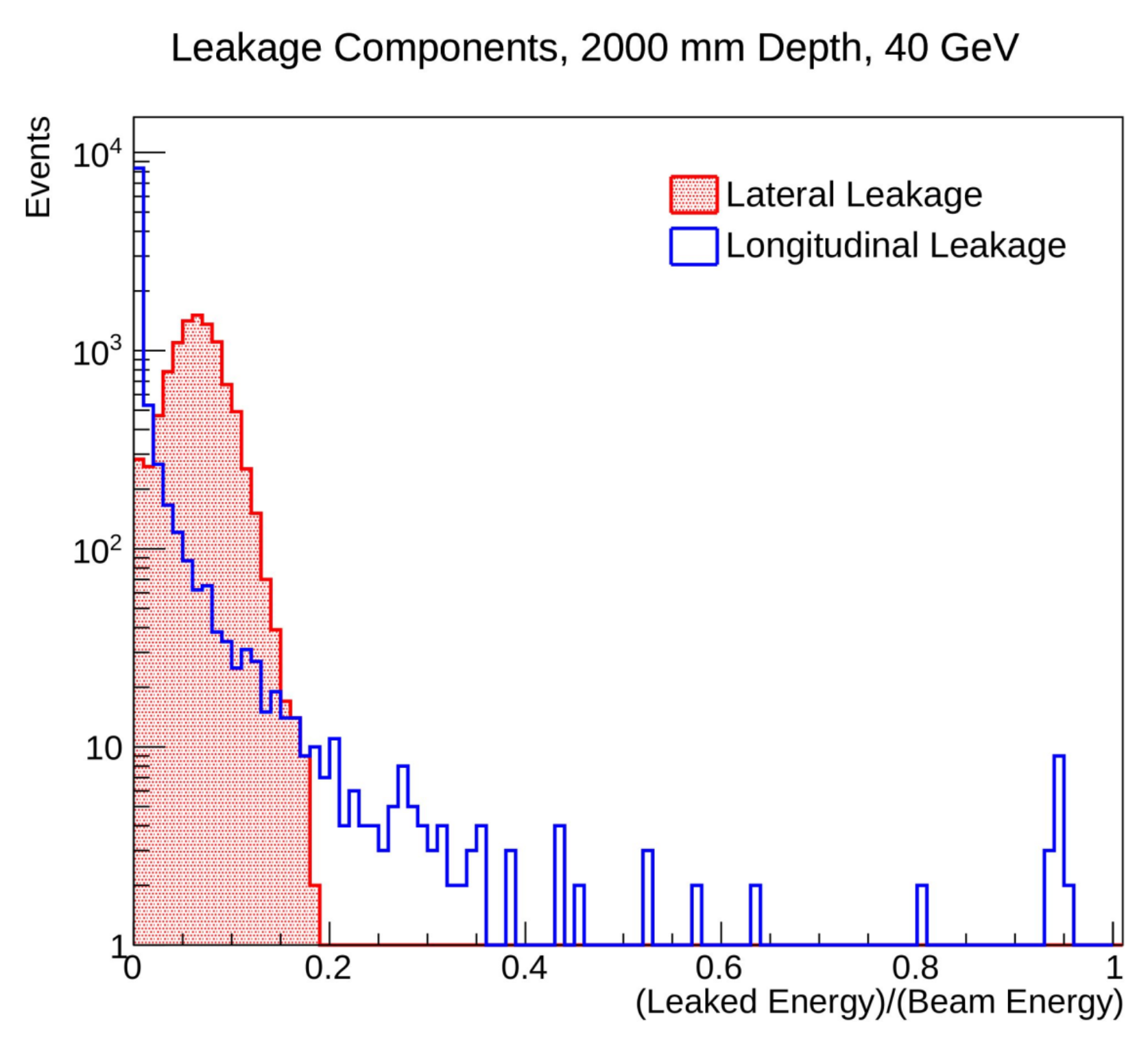}\hfill
    \includegraphics[width=.49\textwidth]{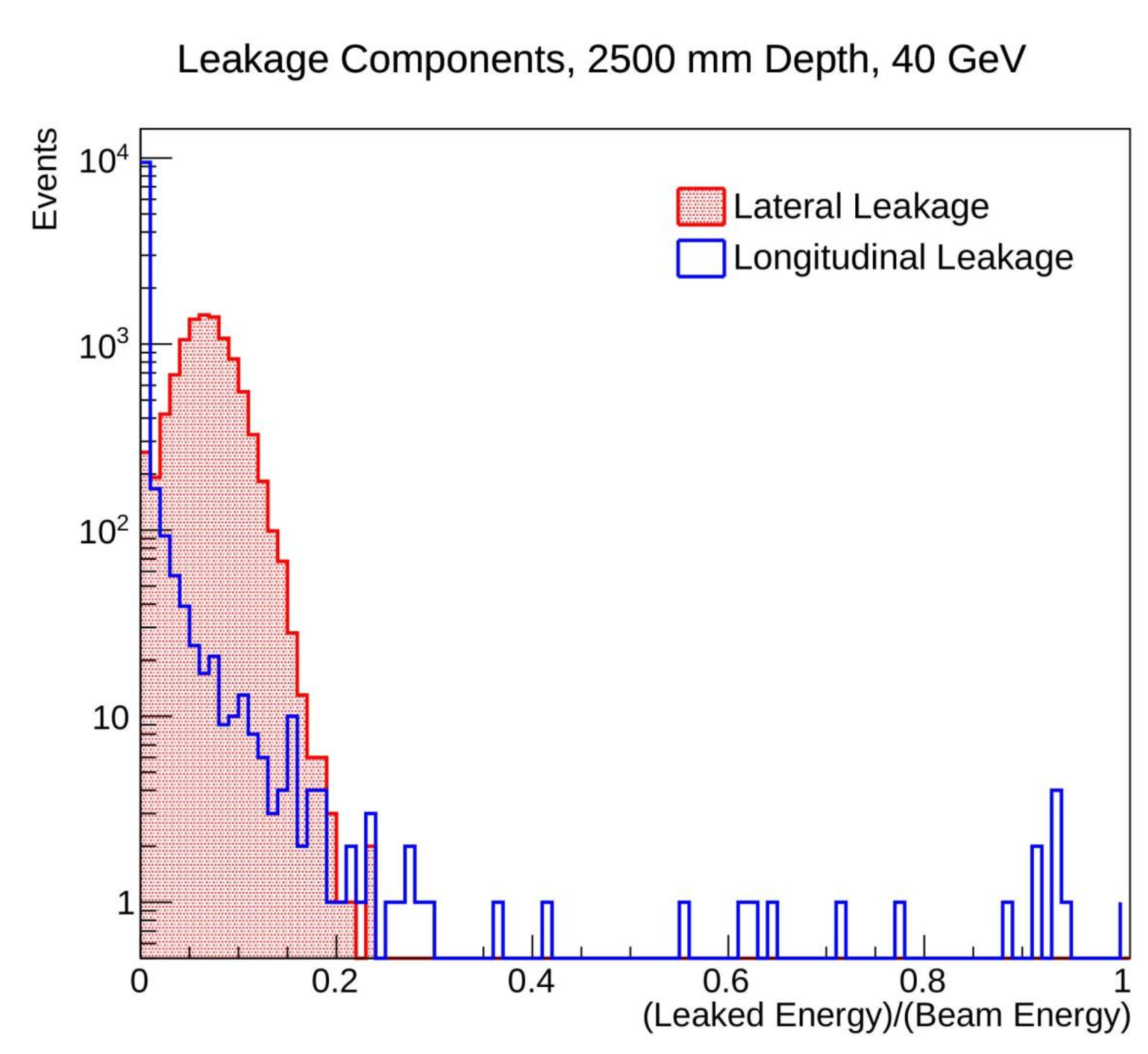}
    \caption{Fraction of the energy lost from the side and the back of the HiDRa prototype separately, for a 2m (left) or 2.5m (right) long calorimeter. A simulation of $10^4$ events with a 40 GeV pion beam impinging on the detector was used.}
    \label{fig:leakage}
\end{figure}

Detailed studies on the larger sources of shower particles leakage outside the prototype volume, with respect to the incoming beam and shower axis direction, were performed through the simulation and are shown in Figure~\ref{fig:leakage}. The largest component of energy losses come from the side of the HiDRa prototype, as should be expected considering the limited dimensions of $65\times65\times 250$ cm$^3$. Because of the not negligible amount of energy not contained inside the active volume, lateral leakage has a considerable impact on the energy resolution. Workarounds to limit this problem, such as adding parts of older RD52 prototypes in the corners of HiDRa, are currently under consideration. The other important component of energy losses is the longitudinal one, that is due to showers initiated considerably deep inside the calorimeter. 
While the yield of lost energy due to longitudinal leakage is significantly lower than the lateral one, a large fraction of energy loss from the back of the detector is associated to a tail of very low reconstructed energy events. The reported energy measurement resolution of the calorimeter comes from the RMS obtained by a Gaussian fit of the reconstructed energy. This results to be almost independent of the few events in the tail of low measured energy events due to high longitudinal leakage. In order to reduce as much as possible this additional source of mis-reconstructed energy events, it was decided to choose a calorimeter depth of 2.5 metres. 

\begin{figure}
    \centering
    \includegraphics[width=.49\textwidth]{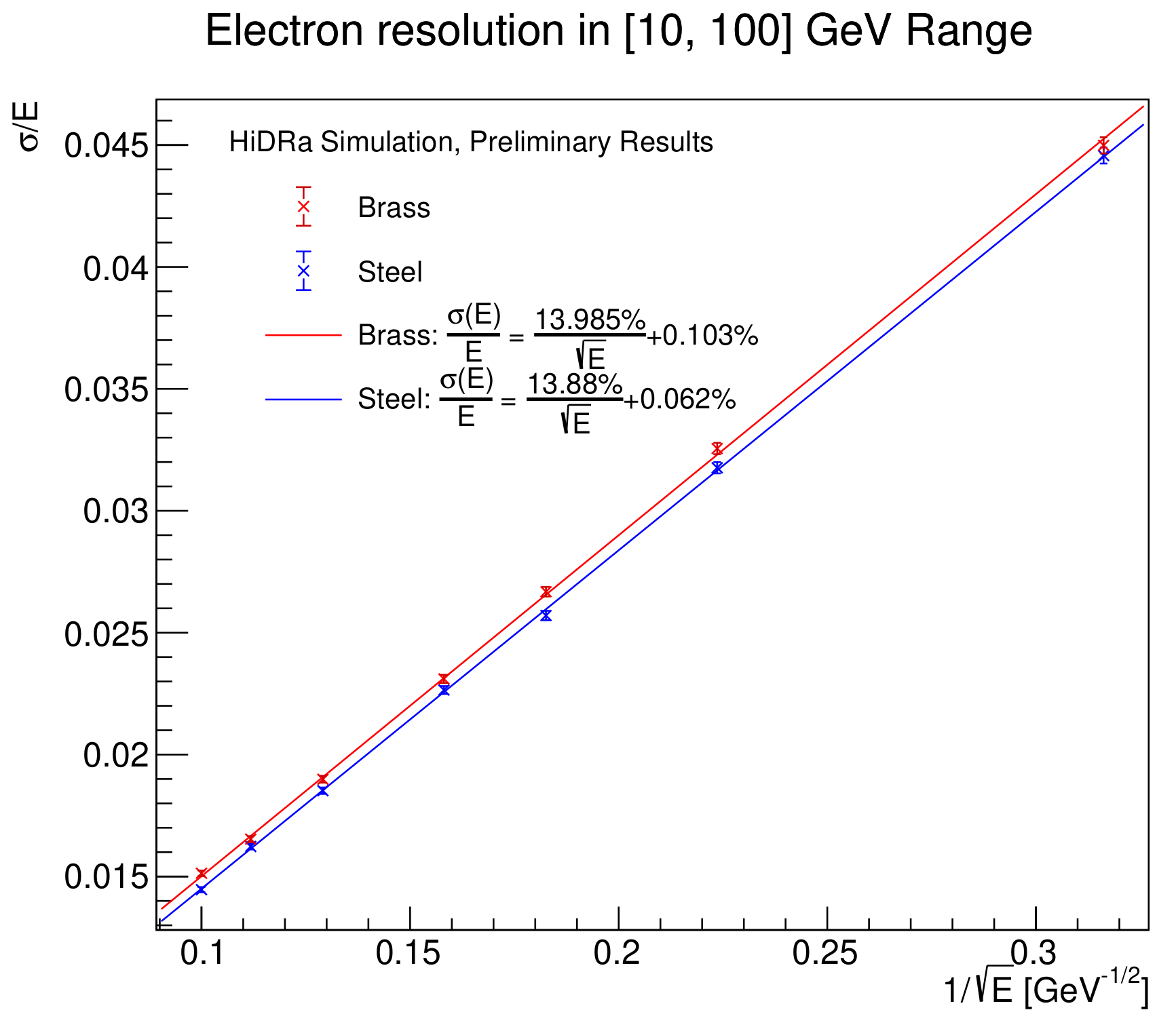}\hfill
    \includegraphics[width=.49\textwidth]{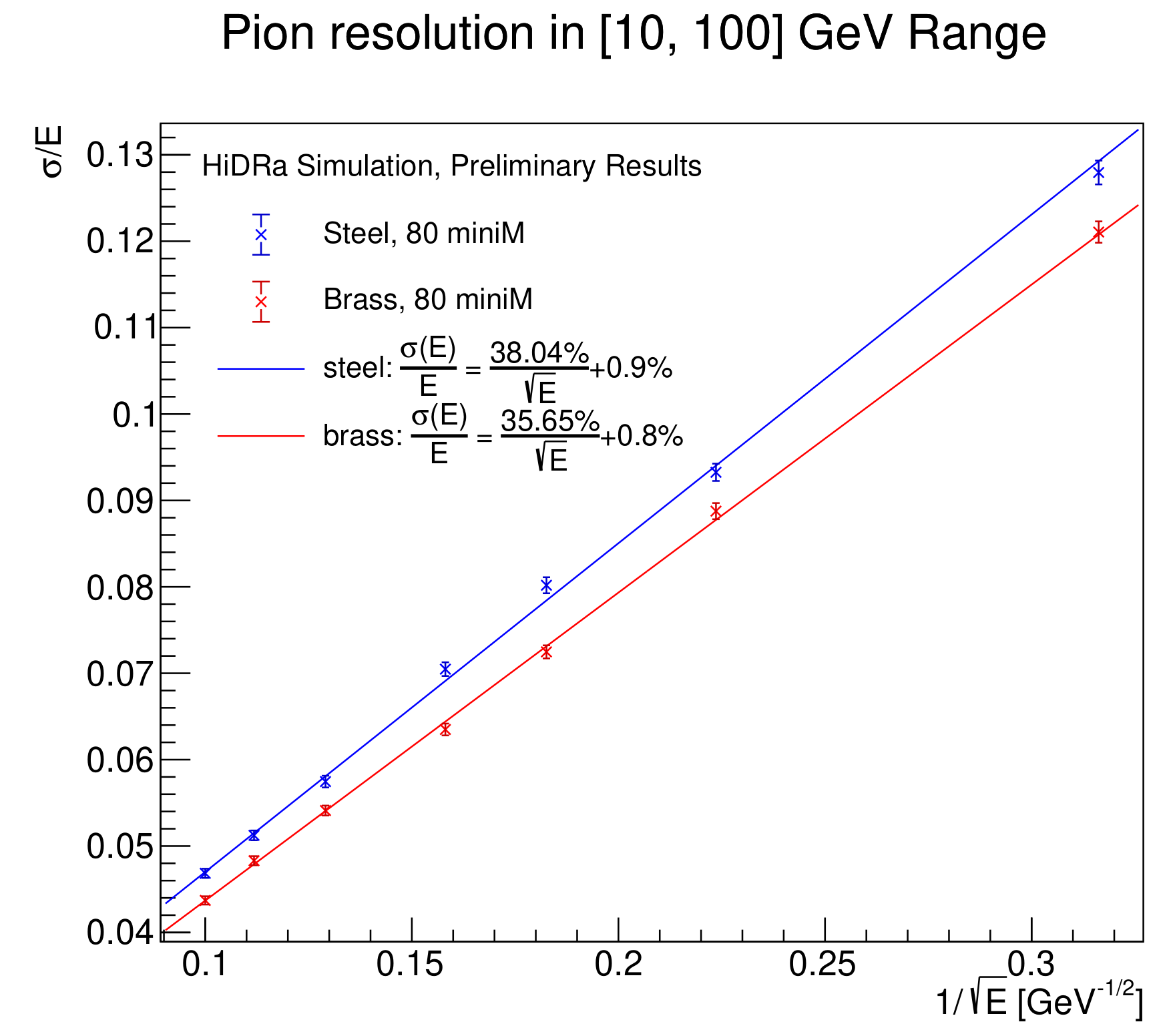}
    \caption{Energy resolution for electrons and hadrons in the range [10, 100] GeV.}
    \label{fig:resolution}
\end{figure}

Figure~\ref{fig:resolution} shows preliminary energy resolution estimates from Geant4 simulations for positrons and single hadrons (positively charged pions) in the range [10, 100] GeV. Two different materials have been considered for the absorber material, stainless steel and the same brass alloy of the previous prototype. For electromagnetic showers the differences between the two materials results almost negligible, while for hadron ones the brass absorber performs slightly better. The improvement in energy resolution has however been considered not large enough to motivate an increased cost of the prototype, and consequently steel was eventually chosen as absorber. In order to estimate the impact of lateral leakage on the detector energy resolution, the modular design was again exploited by simply adding new units in the simulation. With 480 minimodules, a square-like shaped calorimeter with a side of $\sim1.3$ m has been used to estimate the performance the IDEA calorimeter would achieve if it had the same design as the HiDRa one. With an overall containment of $\sim98\%$, stochastic terms of $33.7\%/\sqrt{E}$ and $31.6/\sqrt{E}$ are achieved for steel and brass passive materials, respectively, therefore satisfying the standalone performance required at future colliders.

\section{Conclusions and future prospects}

Calorimetry will play a crucial role at future lepton collider experiments. Different technologies have been proposed to reach excellent performances on hadron jet energy measurements. Among these, highly granular dual-readout calorimeters are promising candidates thanks to the possibility to have both intrinsically linear response to hadrons and the fine segmentation required by modern software techniques. The design of a calorimeter for the IDEA detector at FCC/CEPC steps through the definition of an assembly technique and the assessment of its physics performances. This is carried on with the construction of smaller scale prototypes, whose characterisation at particle beam facilities are also used to validate the simulation from which the IDEA calorimeter features are deduced. The design and current results obtained by two prototypes have been presented: the first one, able to contain electromagnetic showers, demonstrated the feasibility of a fibre calorimeter with SiPM readout. The second one, whose construction is under way, is expected to show optimal results on hadron showers measurements and allow starting with detailed studies on imaging capabilities through its highly granular core.

\bibliographystyle{JHEP}
\bibliography{biblio}





\end{document}